\documentclass[prl,nofootinbib,twocolumn]{revtex4}
\usepackage{graphicx}
\usepackage{bm}
\usepackage{float}
\usepackage{subfigure}
\usepackage{amsmath}
\usepackage{amsfonts}

\def\be{\begin{equation}}
\def\ee{\end{equation}}
\def\bed{\begin{description}}
\def\eed{\end{description}}

\def\bea{\begin{eqnarray}}
\def\eea{\end{eqnarray}}

\def\noi{\noindent}
\def\ba{\begin{array}}
\def\ea{\end{array}}
\def\nn{\nonumber}

\def\vc{v_{\rm c}}
\def\vcmax{v_{\rm c,max}}
\def\rs{r_{\rm s}}
\def\Re{R_{\rm e}}
\def\rhoddm{\rho_{\rm ddm}}
\def\MDDM{M_{\rm ddm}}
\def\Mzero{M_0}

\def\LambdaCDM{\Lambda{\rm CDM}}
\def\ti{t_{\rm i}}
\def\zi{z_{\rm i}}
\def\Msun{M_{\odot}}
\def\Mstar{M_{*}}
\def\rvir{r_{\rm vir}}

\def\fgas{f_{\rm gas}}
\def\r50{r_{50}}
\def\kms{{\rm \,km\,s^{-1}}}

\begin{document}

\title{Secular evolution of galaxies and galaxy clusters in decaying dark 
matter cosmology}

\author{Francesc Ferrer$^1$, Carlo Nipoti$^2$ and Stefano Ettori$^{3,4}$}
\affiliation{
$^1$Physics Department and McDonnell Center for the Space Sciences, Washington University, St Louis, MO 63130, USA\\
$^2$Dipartimento di Astronomia, Universit\`a di Bologna,
via Ranzani 1, I-40127 Bologna, Italy\\
$^3$INAF, Osservatorio Astronomico di Bologna, via Ranzani 1, I-40127 Bologna, Italy\\
$^4$INFN, Sezione di Bologna, viale Berti Pichat 6/2, I-40127 Bologna, Italy
}

\begin{abstract}
  \noi 

If the dark matter sector in the Universe is composed by
metastable particles, 
galaxies and galaxy clusters are expected to undergo
significant secular evolution from high to low redshift.  We show
that the decay of dark matter, with a lifetime compatible with cosmological
constraints, can be at the origin of the observed
evolution of the Tully-Fisher relation of disk galaxies and
alleviate the problem of the size evolution of elliptical galaxies,
while being consistent with the current observational constraints on
the gas fraction of clusters of galaxies.
\end{abstract}

\maketitle

\paragraph{Introduction. }
In the standard $\Lambda$ cold dark matter ($\LambdaCDM$) model
the dynamics of the Universe in the present epoch is dominated
by dark matter (DM) and dark energy.  
The DM, which plays a crucial role in the growth of structure,
does not have appreciable interactions with radiation,
and cannot be in the form of ordinary baryonic matter as deduced from
considerations of big-bang nucleosynthesis together with
observations of the anisotropies in the cosmic microwave background
(CMB). 

The $\LambdaCDM$ model further assumes that the DM particles are
stable, and that any interactions have a negligible effect on the
cosmological evolution.  
This model is extremely successful at reproducing the
matter distribution on Mpc scales and above. However, on the smallest
scales, where galaxies are formed, the simulations of structure growth
seem to disagree with the observations of real
galaxies~\cite{Gilmore:2008}.  Numerical simulations of dark halo
formation in the $\LambdaCDM$ cosmology result in highly cusped
density distributions and abundance of
substructure~\cite{Navarro:2003ew}.
In contrast, the number of satellite dwarf spheroidal galaxies
identified in the local group is lower than the number of subhalos
found in simulations~\cite{Klypin:1999uc}, 
and there is evidence
that the halo of the Milky Way is not cusped at
all~\cite{Binney:2001wu}. 
Also, the recent findings that all dwarf spheroidal galaxies 
in the Milky Way, ranging in
luminosity by more than 4 orders of magnitude, have DM halos with a
common mass ($\sim 10^7 M_\odot$), do not follow from the standard CDM
scenario~\cite{Strigari:2008ib}. At larger scales, recent
high-redshift surveys indicate a structural and kinematical
evolution of galaxies, which has not obvious explanation in the
standard $\LambdaCDM$ structure formation scenario.  In particular,
there is growing evidence that the Tully-Fisher~(TF)~\cite{Tully:1977fu} 
relation for disc galaxies (relating their
stellar mass, $\Mstar$, and rotation speed) evolves with redshift, $z$, in the
sense that higher-$z$ galaxies have higher rotation speeds than
local galaxies of similar $\Mstar$~\cite{Cresci:2009iy,Puech:2009nt}; 
in addition, elliptical galaxies
of given $\Mstar$ are found to be more compact at higher $z$~
\cite{Daddi:2005uv}.

Astrophysical processes, undoubtedly important, could alleviate the
tension between (sub-)galactic structure observations and the
predictions from simulations~\cite{Dekel:1986gu}. In particular,
baryonic processes might affect significantly disc galaxies, which at
high redshift are characterized by high star formation rate
\cite{Cresci:2009iy}. Elliptical galaxies are found to be quiescent at
$z\sim2$, but several mechanisms have been proposed to explain the
size evolution of elliptical galaxies in the context of standard
$\LambdaCDM$~\cite{Bezanson:2009rc}.  However, given the dominant role
played by DM in the growth of structure, a compelling possibility is
that the DM particles are not as cold, stable, or inert as it is
assumed in the $\LambdaCDM$ scenario.  Warmer or strongly
self-interacting DM particles produce more constant density cores with
a higher minimum mass for DM haloes~\cite{Bode:2000gq}, although the
absence of low mass haloes at large redshift would affect the
reionization history~\cite{Barkana:2001gr}.  However, since the
formation of structure at larger, galactic, scales coincides with the
CDM scenario the issues with the fast evolution of early-type galaxies
would remain.

Forfeiting the stability assumption, if the particles composing the DM
sector decay with a lifetime of a few tens of Gyr (consistent with
observations as discussed below), the formation of structure would
proceed until recently, $z \gtrsim 1$, as in the $\LambdaCDM$ model,
successfully reproducing the matter distribution on Mpc scales and
above~\cite{Turner:1984nf}.  Interestingly, the depletion of a
fraction of the DM due to subsequent decays affects the evolution of
structure at galactic and subgalactic
scales~\cite{Turner:1984nf,Cen:2001a,Cen:2001b,SanchezSalcedo:2003pb}.  
Cen~\cite{Cen:2001a}
discussed the possibility that decaying dark matter~(DDM) could solve
the problem of the central cusp in DM halos and the overabundance of
dwarf galaxies, but he also indicated some astrophysical implications
of this scenario on the scale of galaxies and clusters of galaxies
that could be tested observationally.  In this paper, we explore
whether the recent observations of the evolution of disc and
elliptical galaxies from $z \sim 2$, and the available data on the gas
fraction of clusters from $z\sim 1$ can be fit within a DDM scenario
consistent with recent cosmological constraints.

Viable DM candidates with lifetimes on cosmological scales 
frequently appear in 
particle physics models beyond the standard model. 
In fact, much like ${\rm CP}$ is not a symmetry of nature,
there is no fundamental reason 
that requires the discrete symmetries usually
associated with the stability of DM, such as R-parity in 
supersymmetric extensions of the standard model, to hold.  
For instance, in supersymmetric models where a gravitino is
the lightest supersymmetric particle, 
a subset of R-parity violating couplings allows a fast enough
decay of the next-to-lightest-supersymmetric particle 
to avoid conflicts with big-bang nucleosynthesis
predictions, without inducing the decay of the proton at an
unacceptable rate~\cite{Buchmuller:2007ui}. Like in the R-symmetric
case, the required relic density can result from thermal processes,
and, in some scenarios, its decay products could be observed in
cosmic-ray detectors~\cite{Ibarra:2008jk}.
Also, DDM has been recently linked to the origin of neutrino
masses \cite{Lattanzi:2007ux}.
 
\paragraph{Model.}

Let us consider DM particles that decay into relativistic particles
with a characteristic lifetime $\tau$. We will assume that the decay
products are electromagnetically noninteracting so that only the
cosmological evolution is modified with respect to the standard
$\LambdaCDM$ model, through the coupling between the matter and the
radiation sectors.  In particular, observations of the anisotropies in
the CMB put a stringent lower bound on $\tau$, since DDM affects the
evolution of the cosmological perturbations at late times enhancing
the integrated Sachs-Wolfe effect~\cite{Kofman:1986am}.  Using
first year WMAP observations the authors in~\cite{Ichiki:2004vi}
conclude that a particle making the whole of the DM should have a
lifetime $\tau \gtrsim 52$ Gyr (95.4\% C.L.).  A recent study
combining the latest WMAP data, together with type-Ia supernovae,
large scale structure and weak lensing observations, strengthens the
limit to $\tau \gtrsim 100$ Gyr~\cite{DeLopeAmigo:2009dc}. However,
some of these datasets
have been analyzed with the assumption that the 
$\LambdaCDM$ model is correct, which
could bias comparisons between models with a markedly different
cosmological evolution~\cite{Elgaroy:2007bv}.
Moreover, the fits to the CMB datasets assume that the primordial
fluctuations are purely adiabatic, while a sizable admixture of
isocurvature perturbations is allowed by 
observations~\cite{Ferrer:2004nv}. In that
case, the power at low multipoles could be reduced further weakening
the constraints from the integrated Sachs-Wolfe effect. In view of
these facts, we will consider two benchmark models: a DDM model
with $\tau \sim 52$ Gyr, and a more conservative scenario with 
$\tau \sim100$ Gyr.

The decay of DM couples its time evolution with that of radiation (r): 
\bea
\dot{\rho}_{\rm r} + 4 H \rho_{\rm r} &=& \frac{1}{\tau} \rho_{\rm ddm} \nn \\
\dot{\rho}_{\rm ddm} + 3 H \rhoddm &=& -\frac{1}{\tau} \rhoddm,
\label{eq:cont}
\eea while $\rho_{\rm b} \propto a^{-3}$, and $\rho_\Lambda=\rho_\Lambda^0$
as in the $\LambdaCDM$ model, and $ H^2\equiv \dot{a}/a = 8 \pi
  G/3 \left(\rho_{\rm b}+\rho_\Lambda+\rho_{\rm r}+\rhoddm \right)$ is the
expansion rate. 
The superscript $0$ refers to a quantity at present, and $a^0 = 1$.  
Integrating Eq.~(\ref{eq:cont})
we find the look-back time corresponding to a redshift $z$,
$t_0 - t_1 = 1/H_0
\int_{-\log(1+z)}^0{d \eta/\sqrt{\rho/\rho_{crit}^0}}$,
 where $\eta \equiv 1/a = -\log(1+z)$.
To make contact with observations, we also need the luminosity distance, 
$d_L (z) = (1+z)/H_0
\int^{0}_{-\log(1+z)}{ {\rm e}^{-\eta} d \eta /
    \sqrt{\rho/\rho_{crit}^0}}$,
which is related to the angular distance by $d_A = d_L/(1+z)^2$.

In this framework, we want to study the secular evolution of galaxies
and galaxy clusters due to the decay of DM.  In the absence of other
evolutionary effects, from high to low redshift, galaxies should
expand and their rotation speed should diminish, and the gas mass
fraction in clusters should increase, as a consequence of the
variation of mass in DM.  Interestingly, the recent observations of
the evolution of the TF relation for disc galaxies, and the findings
that elliptical galaxies of a given $\Mstar$ are found to be more
compact at higher $z$, appear qualitatively in agreement with this
expected trend.
On the other hand, there is no evidence for an evolution with $z$ of
the gas fraction $\fgas$ of clusters, though --as we show below--
current data cannot exclude a variation of $\fgas$ of the order of
$\sim$20\% at $z\lesssim 1$.

Given that we consider particle lifetimes $\tau$ longer than the age of the
Universe, we can safely assume that structure formation at high $z$ is
unaffected by the decay. Once a DM halo has decoupled from the Hubble
flow and virialized at some time $t_{\rm i}$, its structure and
dynamics will be affected by the DDM density evolving as
\begin{equation}
\rhoddm^{\rm h}(t)=\rhoddm^{\rm h}(\ti)\exp[-(t-\ti)/\tau)].
\end{equation} 
Quantitatively, Cen~\cite{Cen:2001a} considered the case in which
one-half of the DM particles decay into relativistic particles from
the time of halo formation to the present: the depletion of such a
high fraction of DM corresponds to too short a lifetime,
$\tau \lesssim 19$ Gyr. Here, we reconsider the astrophysical
consequences of DDM by assuming lifetimes consistent with recently
determined cosmological constraints, comparing the predictions of
  the DDM model with state-of-the-art observations of galaxies and
  galaxy clusters.

\paragraph{Results.}

\begin{figure}[th]
\includegraphics[width=0.49\textwidth]{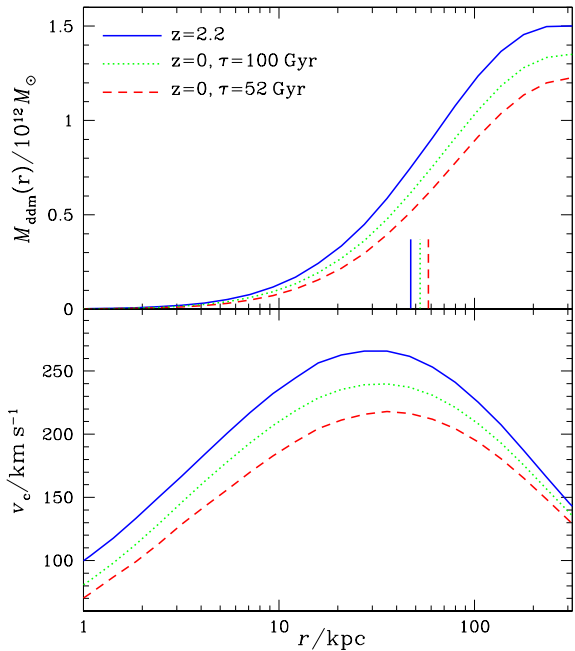}
\caption{\label{fig:vc} Initial ($z=2.2$) and final ($z=0$) DM mass
  enclosed within radius $r$ (upper panel) and circular speed (bottom
  panel) of the two DDM simulations with $\tau=$ 52 and 100 Gyr. The
  vertical lines indicate the initial and final half-mass radius
  $\r50$. The initial distribution is the same in the two cases.}
\end{figure} 

On the scale of galaxies the decay of DM has the effect of an
adiabatic expansion of the DM halo, because $\tau$ is much larger
than the galactic dynamical time~\cite{Cen:2001a}.  To explore
quantitatively the consequences of this process on the structure
and kinematics of galaxies we have performed N-body simulations of
the evolution of the collisionless systems representing the galactic
DM distribution, taking into account the finite lifetime of the DM
particles. We adapted the parallel N-body code
FVFPS~\cite{Nipoti:2003ew}, so that the mass of each DM
particle in the N-body simulation varies in time according to the
exponential decay law: given the collisionless nature of the simulated
system, this is an effective way to model the decrease of the number
of DM particles due to the decay. We assume that at some initial
redshift $\zi$ the DM halo is in equilibrium, with isotropic velocity
distribution and a spherically symmetric Navarro, Frenk, and White
\cite{Navarro:1995iw} density distribution
\begin{equation}
\rhoddm^{\rm h} (r)=\Mzero \frac{\exp\left[ -\left(r/\rvir\right)^2\right]}
{r \left(r+\rs\right)^{2}},
\end{equation}
where $\rs$ is the scale radius, $\Mzero$ is a reference mass, and we
adopt an exponential cutoff to truncate the distribution smoothly at
the virial radius $\rvir$: the total DM mass is $\MDDM=4\pi
\int_0^{\infty}\rhoddm^{\rm h}(r)r^2 dr$. We have performed several
simulations with different choices of the parameters, but here we
present results for a representative pair of simulations with $5\times
10^5$ particles, $\rvir/\rs=10$, $\rs=15$ kpc and
$\MDDM=1.5\times10^{12}\Msun$ starting at redshift $\zi=2.2$ (which we
choose to compare with observations). The two simulations differ only
in the value of $\tau$, which is 52 and 100 Gyr in each case. We have
verified that the N-body system does not evolve significantly in the
absence of DM decay.  In Fig.~\ref{fig:vc} we plot the initial
($z=2.2$) and final ($z=0$) DDM mass profiles (upper panel) and
circular speed ($\vc$) profiles of the simulated halo. In both
simulations the final density distribution is very well represented by
a spherical Navarro, Frenk, and White profile with $\rvir/\rs\simeq 10$. For $\tau=52$ Gyr
the final system has $\sim18\%$ lower DM mass, $23 \%$ larger
half-mass radius $\r50$ and $\sim 22\%$ smaller maximum $\vc$ than the
initial system, while proportionally smaller variations are found in
the case $\tau=100$ Gyr.  Our $\tau=52$ Gyr model at $z=0$ is a
Milky Way like DM halo, with total mass $\sim 1.2\times 10^{12}\Msun$
and $\vcmax\sim216 \kms$. Assuming that most of the galactic stellar
mass is in place at $z= 2.2$ our results suggest that, in the absence
of other evolutionary effects, a $z\sim 2.2$ disk galaxy with stellar
mass similar to that of the Milky Way should have $\sim 22\%$ larger
$\vcmax$ than the Milky Way (the variation in $\vcmax$ might be
smaller if the baryonic mass contributes significantly within the
radius at which $\vc$ peaks). As DDM implies secular evolution for all
galaxies, it predicts that the $\Mstar$-$\vcmax$ relation of galaxies
must evolve with $z$.  Assuming that the maximum circular velocity of
the halo can be taken as a proxy of the maximum line-of-sight rotation
speed of observed disk galaxies, the DDM scenario predicts a higher
normalization of the TF relation of $\sim 22 \% $ in rotation speed at
$z\simeq 2.2 $ than at $z =0$ if $\tau=52$ Gyr.  Remarkably, the
observationally estimated offset between the $z \sim 2.2 $ and the $z
\sim0$ TF relation is $23\%\pm6\%$ \cite{Cresci:2009iy}.

As outlined above, another secular evolution effect predicted by the
DDM model is that galaxies should become more extended from high to
low $z$. It is then tempting to try to explain in this framework the
recent finding that early-type galaxies of similar $\Mstar$ appear to
have smaller half-light radius $\Re$ at high $z$ than at $z=0$, with
an increase in size of at least a factor of $\sim 2-3$ from $z\sim 2$
to the present \cite{Daddi:2005uv}.  To make predictions on the
evolution of $\Re$ in a DDM model, we ran a few simulations similar to
those described above, but in which we allow for the presence of a
stellar component embedded in the DDM halo. We found that $\Re$
increases roughly proportionally to $\r50$, so we can safely use the
results of our DDM-only simulations. Even our DDM with $\tau=52$ Gyr
predicts an increase in size of $\sim 24\%$ and thus cannot explain
the observed dramatic increase in size. As in standard $\LambdaCDM$
cosmology, also in DDM cosmology other effects must be invoked in
order to explain these observations. Nevertheless, it is interesting
that the problem of the size evolution of early-type galaxies is
alleviated in DDM, especially given the possibility that the sizes of
high-$z$ galaxy could be underestimated in current observations
\cite{Hopkins:2009dp}.

\begin{figure}[th]
\includegraphics[width=0.46\textwidth]{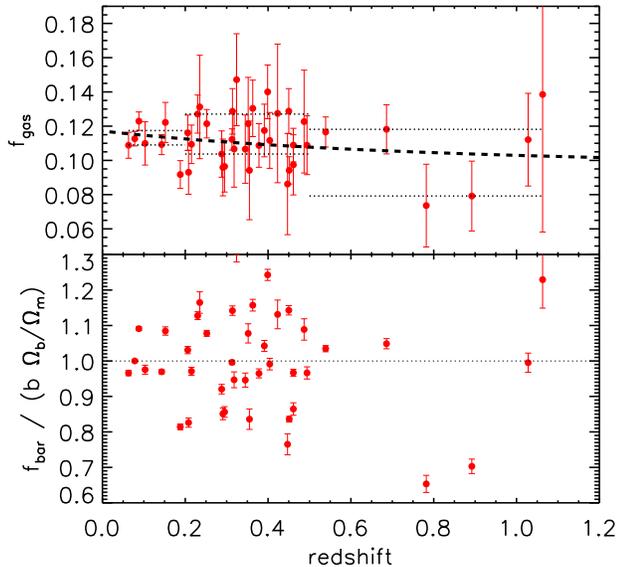}
\caption{{\it Upper panel}: Distributions of the gas mass fraction
  within an overdensity of 2500 with respect to the critical density
  of the Universe at the cluster's redshift in a sample of massive,
  X-ray luminous and relaxed objects from Table~3
  in~\cite{Allen:2008ue}.  The dashed line indicates the expected
  evolution in a DDM scenario normalized to the value that minimizes
  the total $\chi^2$ in the distribution ($\chi^2=41.9$ with 42 data
  points).  The dotted lines show the 1st and 3rd quartile of the
  distribution of the clusters at $z<0.2$, $0.2<z<0.5$, and $z>0.5$.
  {\it Bottom panel}: Distribution of the cluster baryon fraction,
  $f_{\rm bar} = f_{\rm gas} +f_{\rm star}$, with respect to the
  expected cosmic baryonic value in a DDM scenario.  To estimate
  $f_{\rm bar}$, we consider a depletion parameter $b=0.83$ and a
  contribution in stars of $f_{\rm star} = 0.16 h_{70}^{0.5} f_{\rm
    gas}$ (as in~\cite{Allen:2008ue}).  }
\label{fig:fgas}
\end{figure} 

The decay of DM has a straightforward implication for the evolution of
the gas fraction $\fgas$ in clusters of galaxies: in a cluster of
fixed gas mass we expect the gas fraction to evolve in time as
$\fgas\propto \left[ 1 + k \exp(-t/\tau) \right]^{-1}$, with $k$
constant.  In Fig.~\ref{fig:fgas}, we compare this prediction for
$\tau=52$ Gyr with the gas mass fraction measurements
from~\cite{Allen:2008ue}, which represent the largest sample available
with the smallest relative statistical error (median value of about
13\%) and limited intrinsic scatter (the weighted mean scatter around
the best-fit $\Lambda$CDM model is $\sim$7\%).  Considering that all
the dependence on the cosmology can be written as $f_{\rm gas} \propto
d_{L}(z)^{1.5}$~\cite{Allen:2008ue,Ettori:2009wp}, we plot the values,
originally estimated in a $\LambdaCDM$ universe, corrected by the
factor $d_{\rm L, ddm}^{1.5} / d_{{\rm L}, \LambdaCDM}^{1.5}$.  We
show that the observed distribution is consistent with the predicted
variation as function of redshift of about 15\% between $z=1.2$ and
$z=0$.  Moreover, as shown in the bottom panel of Fig.~\ref{fig:fgas},
we obtain that the estimated cluster baryon fraction is well in
agreement with the expected cosmic value $f_{\rm bar} / (b
\Omega_{\rm b, ddm}/\Omega_{\rm m, ddm}) = 0.988 \pm 0.018$
(error-weighted mean and relative error; median: 0.995).
Following~\cite{Allen:2008ue}, here we assume a depletion parameter $b
= f_{\rm bar} / (\Omega_{\rm b}/\Omega_{\rm m})$ within $R_{2500}$
(i.e. the cluster radius encompassing a mean total density 2500 times
the critical density of the Universe at the cluster's redshift) of
0.83. This parameter represents the relative amount of cosmic baryons
that are thermalized within the cluster potential. It can presently
only be inferred from numerical simulations, where both the input
cosmic baryons are known and the amount of them accreting into the
cluster dark matter halo can be traced. When measured over
representative regions of a galaxy cluster, this parameter is
estimated to be lower than unity indicating that a minority of cosmic
baryons are not shock heated during the process of accretion on the
dark matter halo.  However, as discussed in
\cite{Kravtsov:2005,Ettori:2009wp}, the estimate of the depletion
parameter $b$ depends on many aspects of the hydrodynamical
simulations investigated. For example, (i) grid-based, shock-capturing
numerical codes tend to estimate a value $b$ larger by up to 10 percent 
than what is measured in smoothed particle hydrodynamics; (ii)
radiative processes, such as cooling and star formation with feedback
provided from galactic winds, redistribute the baryons within the
virial radius, in particular, in the cluster core, through the
conversion of gas into stars and the replacement of the condensing gas
with gas from the outer regions, with a net effect that reduces the
gas mass fraction and increases slightly the total baryon fraction
with respect to the runs where such processes were not considered;
(iii) the evolution with cosmic time of $b$ is expected to be very
mild and less than 5 percent at $z=1$ with respect to the local
value. To mitigate, at least partially, these uncertainties, observers
prefer to select only hot, massive, and relaxed galaxy clusters that
are expected to be dominated energetically by gravitational collapse
and should be well represented from simulations with no radiative
processes included, as done in the present study by following the
prescription in \cite{Allen:2008ue} (see also discussion on the
estimate of $b$ in \cite{Ettori:2009wp}).

\paragraph{Conclusion. }

We have shown that the current observational constraints on the
secular evolution of galaxies and galaxy clusters are consistent with
a DDM cosmology with a lifetime larger than $\tau \sim 52$ Gyr, which
is the shortest DM decaying timescale that can be reconciled with CMB
observations. This finding is remarkable, because the cosmological
constraints on $\tau$ are completely independent of the set of
observations of galaxies and galaxy clusters at different redshifts
used for comparison, so DM decaying with the shortest timescale
allowed by cosmology could, in principle, give a redshift evolution of
galaxies and galaxy clusters so strong to be excluded by current
observations.  Our results imply that a DDM scenario (with DM lifetime
$\tau \gtrsim 52$ Gyr) is not disfavored with respect to the standard
stable DM scenario, on the basis of astrophysical as well cosmological
arguments.

We find it suggestive that DDM could explain naturally the observed
redshift-evolution of the TF relation of disc galaxies, though this
cannot be considered, {\it per se}, evidence that DM decays. In fact, in our
exploration we isolated the effect of DM decay by neglecting other
processes, such as baryonic physics and galaxy interactions, that
---though poorly constrained--- are expected to influence the
dynamical evolution of disc galaxies.

The gas mass fraction in galaxy clusters can provide a robust test of
DDM models once both the statistical errors and the intrinsic scatter
can be constrained well below the expected change of the order of 15\%
in the redshift range $0-1.2$.  In addition, precise observations of
ultrafaint dwarfs with a wide-field spectrograph such as Gaia or SIM
Lite~\cite{arXiv:0902.3492} will shed light on the DM distribution at
subgalactic scales, where the DDM could also play a distinctive role.

We are grateful to S. R\"as\"anen and U. Sarkar for useful
conversations.  FF and CN acknowledge the hospitality of the Galileo
Galilei Institute where part of this work was completed.  SE
acknowledges the financial contribution from Contract Nos. ASI-INAF
I/023/05/0 and I/088/06/0. FF was partially supported by the NSF
under Contract No.
PHY-0855580 and by the DOE under Contract No. DE-FG02-91ER40628.

\end{document}